\documentclass[12pt,a4paper,final,superscriptaddress]{iopart}
\usepackage[breaklinks=true,colorlinks=true,linkcolor=blue,urlcolor=blue,citecolor=blue]{hyperref}
\providecommand{\onlinecite}[1]{\cite{#1}}

\usepackage{iopams}
\expandafter\let\csname equation*\endcsname\relax
\expandafter\let\csname endequation*\endcsname\relax

\usepackage[utf8]{inputenc}
\usepackage[T1]{fontenc}
\usepackage{amsmath}
\usepackage{amssymb}
\usepackage{printlen}
\usepackage{graphicx}
\graphicspath{{fig/}}
\usepackage{color}
\usepackage{listings}

\definecolor{darkblue}{RGB}{0,0,128}
\definecolor{red}{RGB}{196,0,0}
\definecolor{darkred}{RGB}{160,0,0}
\definecolor{green}{RGB}{0,150,0}
\definecolor{orange}{RGB}{255,64,0}

\newcommand{\dee}[0]{\mathrm d}
\newcommand{\idee}[0]{\,\dee}
\renewcommand{\vec}[1]{\boldsymbol{\mathrm{#1}}}
\newcommand{\X}[0]{\mathrm x}
\newcommand{\cc}[0]{\mathrm c}
\newcommand{\XC}[0]{\mathrm{xc}}

\newcommand{\nl}[0]{\text{nl}}
\newcommand{\Ecnl}[0]{E_{\mathrm c}^{\nl}}

\newcommand{\sect}[1]{Section~\ref{#1}}
\newcommand{\diff}[2]{\frac{\dee #1}{\dee #2}}
\newcommand{\pdiff}[2]{\frac{\partial #1}{\partial #2}}
\newcommand{\fdiff}[2]{\frac{\delta #1}{\delta #2}}
\newcommand{\libvdwxc}{\mbox{libvdwxc}}
\newcommand{\gpaw}{\mbox{\textsc{Gpaw}}}
\newcommand{\octopus}{\mbox{Octopus}}
\newcommand{\quantumespresso}{Quantum \textsc{Espresso}}
\newcommand{\qe}{QE}
\newcommand{\vasp}{\textsc{Vasp}}
\newcommand{\libxc}{\mbox{libxc}}
\newcommand{\norm}[1]{\vert #1 \vert}

\begin{document}

\title{libvdwxc: A library for exchange--correlation functionals in the vdW-DF family}

\newcommand{\upv}{Nano-bio Spectroscopy Group and ETSF Scientific development centre,
Universidad del Pa\'{i}s Vasco UPV/EHU,
%Avenida de Tolosa 72,
%20018 Donostia--San Sebastián
Spain
}
\newcommand{\barcelona}{Departament
de Ci\`{e}ncia de Materials i Qu\'{i}mica F\'{i}sica \& Institut de
Qu\'{i}mica Teòrica i Computacional (IQTCUB), Universitat de Barcelona,
%, c/Mart\'{i} i Franqu\`{e}s 1, 08028 Barcelona,
Spain
}

\newcommand{\chalmersphys}{Department of Physics,
  Chalmers University of Technology, Gothenburg, Sweden}
\newcommand{\jyvaskula}{Department of Chemistry, University of Jyv\"askyl\"a, Jyv\"askyl\"a, Finland}

\newcommand{\yannaffiliation}{YANN.  AFFILIATION.}
\newcommand{\chalmersmicronano}{Department of Microtechnology and Nanoscience,
  Chalmers University of Technology, Gothenburg, Sweden}

\author{Ask Hjorth Larsen$^{1,2,*}$,
  Mikael Kuisma$^{3,4,5,*}$,
  Joakim L\"ofgren$^{3}$,
  Yann Pouillon$^{1}$,
  Paul Erhart$^{3}$
  and
  Per Hyldgaard$^{4}$}

%\author{Ask Hjo Larsen$^{1,2}$,
%  M Kuisma$^{3,4}$,
%  J L\"ofgren$^{3}$,
%  Y Pouillon$^{1}$,
%  P Erhart$^{4}$
%  and
%  P Hyldgaard$^{3}$}

\address{$^1$ \upv}
\address{$^2$ \barcelona}
\address{$^3$ \chalmersphys}
\address{$^4$ \chalmersmicronano}
\address{$^5$ \jyvaskula}
\address{$^*$ Equally contributing authors}
\eads{\mailto{asklarsen@gmail.com}}

%\author{Ask Hjorth Larsen}
%\ead{asklarsen@gmail.com}
%\address{\upv}
%\address{\barcelona}
%\thanks{These authors have contributed equally to this work}
%\affiliation{}
%\affiliation{}
%\author{Mikael Kuisma\normalfont\textsuperscript{*}}
%\address{\chalmersmicronano}
%\address{\jyvaskula}
%\author{Joakim L\"ofgren}
%\address{\chalmersphys}
%\author{Yann Pouillon}
%\address{\yannaffiliation}
%\affiliation{}
%\author{Paul Erhart}
%\address{\chalmersphys}
%\affiliation{}
%\author{Per Hyldgaard}
%\address{\chalmersmicronano}

%\footnote{$\dagger$ Equally contributing authors.}

\begin{abstract}
\noindent
We present \libvdwxc{}, a general library for evaluating the energy and potential for
the family of vdW-DF exchange--correlation functionals.
\libvdwxc{} is written in C and provides
an efficient implementation of the vdW-DF method and can be interfaced
with various general-purpose DFT codes. Currently,
the \gpaw{} and \octopus{} codes implement interfaces to \libvdwxc. The present
implementation emphasizes scalability and parallel performance, and
thereby enables \textit{ab initio} calculations of nanometer-scale
complexes.
The numerical accuracy is benchmarked on the S22 test set whereas
parallel performance is benchmarked on
ligand-protected gold nanoparticles
($\text{Au}_{144}(\text{SC}_{11}\text{NH}_{25})_{60}$) up to 9696 atoms.
\end{abstract}

\maketitle

%unfortunately must use single-column
%\ioptwocol

\section{Introduction}
%\printinunitsof{in}\prntlen{\textwidth}
%\printlength\textwidth

\noindent
Kohn--Sham density functional theory~\cite{HohKoh64, KohSha65} (DFT) is
widely used as a predictive tool in areas ranging from chemistry and
materials physics to biochemical applications.  The continually
increasing computational power is pushing the reach of \emph{ab initio}
modelling to the nanometer scale and provides access to systems with
remarkable diversity.  Many of these are sufficiently sparse in terms
of the spatial distribution of the electron density for dispersive
interactions to play a role~\cite{LanLunCha09}.
%Examples include but are not limited to solvation, biological
%systems, metal organic frameworks, dyadic or triadic photovoltaics
%and nanoparticles.
In addition to requiring extensive computational resources, the common
denominator among these systems is the importance of van der Waals
(vdW) interactions.

This situation has prompted the development of the vdW
density functional (vdW-DF)
method~\cite{BerCooLee15,Thonhauser_2015:spin_signature}
which comprises a growing family of exchange--correlation (XC) functionals.  These include
vdW-DF1~\cite{DioRydSch04,Dion05},
vdW-DF2~\cite{LeeMurKon10},
vdW-DF-C09~\cite{cooper10p161104},
vdW-DF-cx~\cite{BerHyl14},
vdW-DF-optb86~\cite{KliBowMic11},
rev-vdW-DF2~\cite{hamada14},
vdW-DF-optPBE,
vdW-DF-optB88~\cite{KliBowMic10},
BEEF-vdW~\cite{PhysRevB.85.235149}, % the Oxford comma
and mBEEF-vdW~\cite{PhysRevB.93.235162}.
In those functionals the nonlocal correlation forces are captured through a
formal analysis of screened response in the electron gas~\cite{BerCooLee15}.
Closely related is also VV10~\cite{vv10}
and revised VV10~\cite{PhysRevB.87.041108},
which adapt the vdW-DF framework to a simpler response description.
The vdW-DF method can characterize both pure vdW forces and vdW forces
in combination with other types of binding~\cite{BerArtCoo14,BerCooLee15}.
The vdW-DF method has gained a reputation for providing
accurate characterizations and predictions for a range of systems
including but not limited to
metals~\cite{KliBowMic10, GhaErhHyl17},
layered materials~\cite{Bjo14, ErhHylLin15, LinErh16, BerArtCoo14, MurSunIma15}, and
molecular systems~\cite{BerHyl14, LofGroMot16, KuiLunMot16a, RanBerSha16, BroRanNea16}.

%This situation has prompted the development of a family of so-called
%van der Waals density functionals (vdW-DF), which include vdW-DF1
%\cite{DioRydSch04}, vdW-DF2 \cite{LeeMurKon10}, vdW-DF-cx
%\cite{BerHyl14}, as well as vdW-DF-optPBE and vdW-DF-optB88
%\cite{KliBowMic10}. Although the presented functionals comprise van
%% XXXX other vdW functionals?
%der Waals interactions and are named accordingly, we wish to stress
%that first and foremost they can be considered as a general class of
%exchange-correlation (XC) energy approximations of a particular
%functional form with predictive power for diverse classes of materials
%ranging from metals \cite{GhaErhHyl16}, layered materials
%\cite{BerHyl14, ErhHylLin15, LinErh16} to molecular systems.

% AHL: taking liberty to remove this sentence:
%The vdW-DF family is a relatively new class of exchange--correlation (XC)
%functionals, and attention should be given to seek optimal implementation
%strategies.
In the vdW-DF family, the XC energy can be written in the form
\begin{align}
  E_\text{xc}[n]
  &= E_\text{x}^\text{GGA}[n] +
  E_\text{c}^\text{LDA}[n] + E_\text{c}^\text{nl}[n].
  \label{eq:vdw-df-energy}
\end{align}
Here, $E_\text{x}^\text{GGA}[n]$ is an exchange energy functional in the
generalized gradient approximation (GGA), and $E_\text{c}^\text{LDA}[n]$ is the
well-established Perdew--Wang parametrization of the correlation energy in the
local-density approximation (LDA)~\cite{PerWan92}.
The third term is the characteristic non-local correlation energy
functional of the vdW-DF method that describes the vdW interactions,
\begin{align}
  \Ecnl[n]
  &= \frac 12 \iint n(\vec r) \phi\big(q_0(\vec r), q_0(\vec r'), \norm{\vec r - \vec r'} \big)  n(\vec r') \idee \vec r \idee \vec r',
  \label{eq:nonlocal}
\end{align}
which has a different formal structure from semilocal functionals.
In \eqref{eq:nonlocal},
$\phi(q_0, q_0', r)$ is the vdW kernel,
and $q_0(\vec r)$ represents an inverse length scale that characterizes the
roll-over in the vdW-DF plasmon-pole description.
This roll-over function $q_0(\vec r)$
is determined by the energy per particle of an internal GGA-type
functional~\cite{LeeMurKon10,HylBerSch14,BerCooLee15}, and is therefore \textit{local},
depending exclusively on the local density and its
gradient.

The integral \eqref{eq:nonlocal} is six-dimensional,
but it can
be efficiently computed using a spline decomposition
of the two spatial variables
together with the Fourier convolution theorem~\cite{RomSol09}.

While the vdW-DF method has been implemented, e.g., in the
\textsc{Siesta}~\cite{0953-8984-14-11-302},
\textsc{Abinit}~\cite{Gonze20092582},
\quantumespresso{}~\cite{GiaBarBon09},
\vasp{}~\cite{KreFur96a},
and \gpaw{}~\cite{EnkRosMor10}
codes, both
performance and results vary. Typical vdW forces are small,
corresponding to shallow potential energy landscapes with weak
curvature. Bonding distances are sensitive to even small
differences in implementations.  Hence a standardized reference
implementation is of great interest.

We therefore
\footnote{One reason to start work on \libvdwxc{} was
the desire to include vdW-DF in the
\octopus{} code~\cite{octopus_review} and improve the
performance of the existing vdW-DF implementation
in \gpaw{}~\cite{Wellendorff2010,EnkRosMor10}
without further increasing the number of
of separate implementations.}
present here the \libvdwxc{} software library that
provides an efficient implementation of the vdW-DF method and can be
interfaced with various general-purpose DFT codes.

% AHL: Removed repeat text from abstract.
%Our implementation
%emphasizes scalability and parallel performance, and thereby enables
%\textit{ab initio} calculations of nanometer-scale complexes.
%In
%particular it should not be necessary for users to invest significant
%thought in computational performance, just as in the case of regular
%GGAs.
What \libvdwxc{} does is to compute the non-local
correlation energy \eqref{eq:nonlocal}
and its derivatives for evaluating the potential.
The corresponding semilocal functionals
in \eqref{eq:vdw-df-energy}
are available from the \libxc{} library~\cite{MarOliBur12}.
Indeed the success of \libxc{}
has been a major source of inspiration for \libvdwxc.
\libvdwxc{} complements \libxc{} so that
all the functionals in the vdW-DF
family mentioned above are accessible given the two libraries.
The reason why \libvdwxc{} is a separate library
and not part of \libxc{} is the big difference
between semilocal and fully non-local functionals.
\libxc{} relies on the fact that all the supported functionals
are evaluated pointwise: At each point, the calculation requires
only the density and its derivatives in that point.
Since the vdW-DF functionals are non-local,
\libvdwxc{} must instead work on the full density.
The evaluation also requires separate array
allocations and fast Fourier transforms (FFTs), and is non-trivial to
parallelize.  \libvdwxc{} therefore depends on FFTs from the FFTW
library~\cite{FFTW05} or optionally PFFT~\cite{Pi13}.
Parallelism is supported through MPI.

%of the vdW-DF family
%of functionals, which follow the general form defined by
%Eq.~\eqref{eq:nonlocal}.
%The existing \libxc{} library
%of semilocal functionals
%already provides the corresponding exchange functionals
%\cite{MarOliBur12}.

%For physisorption and weak binding, local and semi-local density functionals are not well suited for these problems.
%However, there a general trend in the vdW-DF
%fuctional family is becoming a generally applicable
%functional\cite{BerHyl14,KliBowMic11} (bulk properties,
%surfaces, clusters, dispersive interactions) instead being profiled as
%the functional for weak dispersive interactions.

%In short, the complexity of vdW-DF compared to
%semilocal functionals goes against
%the grain of the interface of \libxc{}.
%As a result, implementing
%vdW-DF within \libxc{} would not provide any technical advantage.

%By optimizing memory, cache and intra-process
%communications, our implementation surpasses the existing
%\gpaw{} code by two orders of magnitude on extremely large
%systems, and thanks to its library form enables similar performance
%also for other codes. This implies that the computational effort for
%vdW-DFs is comparable to conventional semi-local XC functionals
%independent of system size enabling large scale calculations.

\libvdwxc{} does not at present include
other vdW-inclusive approaches such as the Wannier function
approach~\cite{Sil08, Silvestrelli09p5224}, the
Tkatchenko--Scheffler method~\cite{TkaSch09}, the Grimme
D-correction series~\cite{Gri06, GriEhrGoe11},
or exchange-hole dipole moment theory~\cite{doi:10.1063/1.1884601}.

The remainder of this article is organized as follows.  In the
following section, we review the Rom\'{a}n-P\'{e}rez--Soler algorithm.
(Background information concerning construction and interpretation of
the vdW-DF method can be found in Refs.~\onlinecite{BerCooLee15} and
\onlinecite{HylBerSch14}). \sect{sect:parallelization} is a detailed
description of the parallel implementation of vdW-DF.
Calculated results are
benchmarked against other codes in
\sect{sect:benchmarks}. We demonstrate the excellent
scalability of the library in \sect{sect:performance}.
Finally, a brief technical description is provided in
Section \ref{sec:thelibrary}.

\section{Rom\'{a}n-P\'{e}rez--Soler method}
\label{sect:background}

%The van der Waals energy functional consists of a semilocal exchange
%and correlation functional plus a non-local contribution to the
%correlation, where the latter takes the form of the double integral
%\eqref{eq:nonlocal}.
%Each pair of points $\vec r$ and
%$\vec r'$ contributes an energy determined by the density semilocally
%at the two points $\vec r$ and $\vec r'$ and the distance
%$\norm{\vec r - \vec r'}$ between them.

\noindent
While Eq.~\eqref{eq:nonlocal} is prohibitively expensive to
evaluate by direct numerical integration in six dimensions
\footnote{
However, on sufficiently large systems, theoretical $\mathcal O(N)$ scaling may be achieved
using purely real space methods due to the real-space cut off of the kernel.
\cite{Berland20111800}
},
it can be efficiently
approximated by the 2D interpolation method by Rom\'{a}n-P\'{e}rez and
Soler~\cite{RomSol09}.
The idea is to replace the continuous $q_0$ parameters of the
kernel $\phi(q_0, q_0', r)$
by a grid of $M$ discrete $q_0$ values, such that the kernel is
instead
described by $M\times M$ radial functions $\phi_{\alpha\beta}(r)$
with $\alpha, \beta = 0, 1, \ldots, M - 1$.
A spline representation over the $q_0$ grid is used
to retain high numerical precision between the points of the discrete $q_0$
grid~\cite{RomSol09}.
Specifically this is done by introducing
$M$ helper splines $p_\alpha(q_0)$ such that
%In this method, the two $q_0$ variables of the kernel are discretized to a
%grid of $M$ values of $q_0$ (typically $M = 20$).
$p_\alpha(q_0^\beta)=\delta_{\alpha\beta}$:
Each function takes the value 1 on its ``own'' mesh point and 0 on all the
others.  Between the mesh points the functions
oscillate weakly.
The spline representation interpolates
both $q_0$ and $q_0'$ of $\phi(q_0, q_0', r)$,
%with each pair $(\alpha, \beta)$ corresponding
%to a particular $(q_0^\alpha, q_0^\beta)$ to
combining to
produce an overall accurate
and smooth representation over the whole range of $(q_0, q_0')$.
The full definition of the helper
function $p_\alpha(q_0)$
at the grid point $q_0^\beta$ plus a small displacement
$\dee q_0$ is
\begin{align}
  p_{\alpha}(q_0^\beta + \dee q_0) = \sum_{c=0}^{3} a^c_{\alpha\beta} (\dee q_0)^c, 0 \leq {\dee q_0} < q_0^{\beta + 1}-q_0^\beta.
\end{align}
The coefficients $a_{\alpha\beta}$ are determined by the aforementioned
condition that $p_\alpha(q_0^\beta)=\delta_{\alpha\beta}$, along with the requirement
of continuity of derivatives up to second order.
It was found in Ref.~\onlinecite{RomSol09} that $M=20$ mesh points
were sufficient to achieve good precision.  This remains the standard procedure
today, although a smoother representation of the kernel has recently
been suggested which reduces the required number of
points~\cite{doi:10.1063/1.4832141}.

We now define the auxiliary quantity
\begin{align}
\theta_\alpha(\vec{r}) = n(\vec r) p_\alpha(q_0(\vec r)),
\end{align}
which is the key quantity in actual computations.
The energy can then be written as
\begin{align}
  \Ecnl[n] &= \frac12 \sum_{\alpha\beta} \iint  \theta_\alpha(\vec r)
  \phi_{\alpha\beta}(\norm{ \vec r' - \vec r }) \theta_\beta(\vec r')
  \idee \vec r' \idee \vec r \nonumber\\
  &= \frac12 \sum_\alpha \int
  \theta_\alpha(\vec r) F_\alpha(\vec r) \idee \vec r,
\end{align}
where
\begin{align}
  F_\alpha(\vec r) = \sum_\beta \int
  \phi_{\alpha\beta}(\norm{\vec r - \vec r'}) \theta_\beta(\vec r')
  \idee \vec r'.
\end{align}
This integral is a convolution, and the energy
can therefore be written using the convolution theorem as
\begin{equation}
\Ecnl[n] = \frac12 \sum_{\alpha\beta}\int
\theta^*_\alpha(\vec k) \phi_{\alpha \beta}(k) \theta_{\beta}(\vec k) \idee \vec k,
\label{enlk}
\end{equation}
where $\theta_\alpha(\vec k)$ is the Fourier transform of
$\theta_\alpha(\vec r)$.

Practical calculations also require the potential,
which is defined as the derivative of the energy with respect to the density,
\begin{align}
  &v_{\mathrm c}^\nl(\vec r)
  \equiv
  \fdiff{\Ecnl[n]}{n(\vec r)}
  = \left.\pdiff{\Ecnl[n]}{n(\vec r)}\right|_{\sigma}
  + \left.\pdiff{\Ecnl[n]}{\sigma(\vec r)}\right|_{n} \diff{\sigma(\vec r)}{n(\vec r)},
\end{align}
where $\sigma(\vec r)=\norm{\vec \nabla n(\vec r)}^2$.
The partial derivatives are
\begin{align}
  \pdiff{\Ecnl[n]}{n(\vec r)}
  &=\sum_\alpha F_\alpha(\vec r) \left[
    p_\alpha(q_0(\vec r)) + n(\vec r) p_\alpha'(q_0(\vec r))
    \pdiff{q_0(\vec r)}{n(\vec r)}
    \right],\label{eq:dedn}\\
  \pdiff{\Ecnl[n]}{\sigma(\vec r)}
  &= n(\vec r) \sum_\alpha F_\alpha(\vec r)
  \diff{p_\alpha(q_0(\vec r))}{q_0(\vec r)}
  \pdiff{q_0(\vec r)}{\sigma(\vec r)}.\label{eq:dedsigma}
\end{align}
\libvdwxc{} implements these partial derivatives, while the calling DFT code
is responsible for calculating the density-derivative $\sigma(\vec r)$
and combining the calculated partial derivatives \eqref{eq:dedn}
and \eqref{eq:dedsigma} to obtain the potential.
Any DFT code that supports GGAs already implements the requisite
functionality, which is also the requirement for calling \libxc.
For completeness we provide the expression for $q_0(\vec r)$ in the
appendix (see also Ref.~\onlinecite{DioRydSch04}).

\libvdwxc{} currently uses the standard kernel and mesh representation
from \quantumespresso.  This means using the same 20 mesh points $q_0^\alpha$ and the
same $20 \times 20$ radial functions $\phi_{\alpha\beta}(k)$;
Support for multiple pluggable kernels (e.g.~the GPAW
parametrization) is under development.

\section{Computation and parallelization}
\label{sect:parallelization}

\begin{table}
  \centering
  \begin{tabular}{cccc}
    Expression & Description\ & Scaling & Expensive \\
    \hline\hline
    %$q_0\big(n(\vec r), \norm{\nabla n(\vec r)}^2 \big)$
    $q_0\big(n(\vec r), \sigma(\vec r) \big)$
    & math & $\mathcal O(N)$ & no \\
    %$p_\alpha(q_0(\vec r))$ & Splines & $\mathcal O(N)$ & no
    % = n(\vec r) p_\alpha(q_0(\vec r))
    $\theta_\alpha(\vec r) $ & splines & $\mathcal O(N)$ & no\\
    $\theta_\alpha(\vec k)$ & 20 $\times$ FFT & $\mathcal O(N \log N)$ & yes\\
    $\Ecnl$, $F_\alpha(\vec k)$ & 400 integrals & $\mathcal O(N)$ & yes\\
    $F_\alpha(\vec r)$ & 20 $\times$ iFFT & $\mathcal O(N \log N)$ & yes\\
    $v_{\mathrm c}^\nl(\vec r)$ & math & $\mathcal O(N)$ & no
  \end{tabular}
  \caption{Steps in a vdW calculation and their computational scaling with respect to the total number $N$ of real-space points.}
  \label{tab:vdw-computation-steps}
\end{table}

%The energy is evaluated efficiency as a convolution in Fourier space
%\begin{align}
%  \Ecnl = \frac12 [\Omega?] \sum_{\alpha\beta}
%\theta_\alpha^*(\vec k) \phi_{\alpha\beta}(k) \theta_\beta(\vec k)
%\idee \vec k,
%\end{align}
%where $\theta_\alpha(\vec k)$ is the Fourier transform of
%\begin{align}
%  \theta_\alpha(\vec r) = n(\vec r) p_\alpha(q_0(\vec r)),
%\end{align}
%with
%\begin{align}
%  q_0(\vec r) = \textrm{[definition of q0]}
%\end{align}

%\input{math}

\noindent vdW calculations are significantly more complex than ordinary GGA
calculations. The calculation needs to allocate at least 20 functions
($\theta_\alpha(\vec r)$) on top of a GGA calculation.  Furthermore, vdW
systems are often very sparse, and due to the long range of the vdW
interactions, a larger vacuum region is often required to
avoid artificial interactions between periodic images.  Good
performance is therefore important and in particular good parallel
scalability.
The computational complexity of a vdW calculation is
$\mathcal O(N \log N)$ due to the Fourier transforms, whereas standard
Kohn--Sham DFT scales as $\mathcal O(N^3)$.  Therefore the vdW
calculation will hardly topple the computational budget for large systems
unless it is grossly unscalable.
Since part of the reason to write \libvdwxc{} is to provide a
scalable implementation, we discuss the
technical aspects in some detail below.

%While the vdW
%calculation is and remains more expensive than standard GGAs, what
%What matters to performance in practice is whether
%the computation time and memory requirements are low compared
%to the full DFT calculation.  Ideally it should not be any more troublesome
%than any GGA, but the allocation of 20 large arrays ($\theta_\alpha(\vec r)$)

%Many vdW systems are very sparse, though, so a large simulation cell
%is often necessary compared to the actual number of electrons.
%Also since the vdW interactions are long-ranged, it may be necessary
%to include more surrounding space around the system to avoid artificial
%interactions between periodic images.

%Therefore vdW calculations typically must deal with 20 quite large arrays
%($\theta_\alpha(\vec r)$) on top of what a GGA does.

The calculation of vdW energy and potential involves the
steps listed in Table~\ref{tab:vdw-computation-steps}.
The computationally important parts of the calculation are the Fourier
transforms, and to a lesser extent the convolution in Fourier space for
calculating $F_\alpha(\vec k)$ and the energy.

The ($\alpha$, $\vec r$) and ($\alpha$, $\vec k$) arrays can be distributed
over $\alpha$ or $\vec r/\vec k$.
Parallelization over $\alpha$ is not overly promising, though, as there are only $M=20$ of them.
\libvdwxc{} therefore distributes $\theta_\alpha(\vec r)$ over $\vec r$
and, taking advantage of the parallel features of FFTW,
correspondingly distributes $\theta_\alpha(\vec k)$ over $\vec k$.
We have found this type of parallelization to be quite sufficient
(see Section \ref{sect:performance}),
and so have not implemented simultaneous parallelization over $\alpha$.
%It is possible that further performance improvements could be achieved from parallelizing
%\emph{additionally} over space and $\alpha$ but we find the Fourier transforms
%to parallelize sufficiently enough without this additional complexity
%(see Section \ref{sect:performance}).

%In the end,
%questions of efficiency revolve mostly
%around the FFTs.
\libvdwxc{} assumes that the
electron density is provided on a uniform real-space grid within a
simulation cell that is
a parallelepiped.
This means $\theta_\alpha(\vec k)$ is calculated from 20
standard 3D FFTs of the real-valued functions $\theta_\alpha(\vec r)$.
A 3D FFT on a $p \times q \times r$ grid consists of many 1D FFTs:
First $p \times q$ FFTs of length $r$ over the $z$ axis,
then likewise over $y$ and $x$.
%the result of which is then transformed likewise over
%$y$ and $z$.
Distributing the many 1D FFTs over many cores is more efficient than
performing inherently parallel 1D FFTs.
%, but requires parallel transposes.
In FFTW the data is initially distributed along the $x$
(least memory-contiguous) axis.
The FFTs over $z$ and $y$ can be performed immediately, but
then the data must be redistributed using a parallel transpose
to allow the FFT over $x$.  This is handled by FFTW, but the initial
data distribution must still be set up by the caller.
%over $z$ and $y$. Then a parallel transpose of all the data
%is required to allow the FFT over $x$, after which the data is
%distributed over, e.g., $y$.

%In FFTW the
%data must be initially distributed along $x$, which in this context
%means the least memory-contiguous axis.
Codes that wish to
interface with \libvdwxc{} must therefore redistribute their
density to the initial FFTW-MPI layout:
Process $n$ is responsible for $B$ elements along the $x$ axis numbered
$nB$ to $(n+1)B$ (truncated if necessary to total number of points),
where the blocksize $B$ should divide
all the points between all available cores as evenly as possible.
Implementing this redistribution is the most complex task
from the perspective of the calling DFT code, although many codes
may already
support this feature if they employ parallel FFTs
(e.g.~Octopus~\cite{JCC:JCC23487}).

The distribution takes advantage of as many cores as there
are grid points along the coordinate axes. If more cores are available, they
will be idle during this stage.
This does not occur in normal calculations, but is insufficient
for massively
%This is sufficient for most applications, but not for massively
parallel architectures.  For this case \libvdwxc{} can use
PFFT~\cite{Pi13}, which relies on FFTW while extending parallelization to two dimensions
(PFFT in general supports parallelization over $n-1$ dimensions
for an $n$-dimensional Fourier transform).
This is much more scalable, but requires another parallel transpose between
the transforms over $z$ and $y$.
For common forms of parallelism we therefore recommend standard FFTW with MPI.

\newcommand{\arr}[1]{\bar{#1}}

The full calculation procedure is:
\begin{enumerate}
\item Evaluate the functions $\theta_\alpha(\vec r)$
  as an array $\arr \theta_{x y z \alpha}$ distributed over $x$.
  Here and below, bars ``$\arr{\ }$'' denote arrays
  indexed by one or more subscripted quantities.
  The \emph{rightmost} subscripted quantity
  (in this case $\alpha$) is contiguous in memory.
\item Compute the FFT of $\arr \theta_{x y z \alpha}$ \emph{in-place}
  (input and output buffers are the same).
  This yields $\arr \theta_{k_x k_y k_z \alpha}$,
  now distributed over $k_y$.
\item For each $\vec k=(k_x, k_y, k_z)$, processing only the local $k_y$ for
  each process:
  \begin{enumerate}
  \item Compute the kernel as an
    $M \times M$ matrix $\arr \phi^k_{\alpha \beta}$ using linear interpolation
    to resolve the kernel function values for
    the continuous variable $k$ from the discrete $k$-grid on which they
    are represented.
  \item Calculate $F_\alpha(\vec k)$ as the matrix--vector product
    \begin{align}
      \arr F_\alpha^{\vec k} = \sum_\beta \arr \phi^k_{\alpha \beta} \arr \theta_{k_x k_y k_z \beta}.
    \end{align}
    $\arr F_\alpha^{\vec k}$ is a buffer of size $M$.
    %(Note that parallelization over $\alpha$ would inconveniently
    %make these matrix products parallel.)
  \item Sum up the energy contributions
    \begin{align}
      \Delta \Ecnl = \sum_\alpha \arr \theta_{k_x k_y k_z \alpha}^* \arr F_\alpha^{\vec k},
    \end{align}
    and write the buffer $\arr F_\alpha^{\vec k}$ back into the same buffer as
    $\arr \theta_{k_x k_y k_z \alpha}$, now denoted $\arr F_{k_x k_y k_z \alpha}$.
  \end{enumerate}
\item Compute the inverse in-place FFT of $\arr F_{k_x k_y k_z \alpha}$ to obtain
  $\arr F_{x y z \alpha}$, once again distributed over $x$.
\item Compute the energy-derivatives using \eqref{eq:dedn} and
  \eqref{eq:dedsigma}.
\end{enumerate}

\noindent
The above procedure follows the efficiency recommendations of FFTW:
\begin{enumerate}
\item
  The memory layout of the input buffer is \emph{strided},
so for each point $\vec r$ of space, values for the $M=20$ different $\alpha$
are stored contiguously rather than the other way around,
\item
  the transform is performed \emph{in-place}, and
\item
  the output array remains in its \emph{transposed} form, i.e., it is distributed over the second axis
  instead of the first.
\end{enumerate}
We note that only a single work buffer is used for the four quantities
$\theta_\alpha(\vec r)$, $\theta_\alpha(\vec k)$,
$F_\alpha(\vec k)$, and $F_\alpha(\vec r)$.
\libvdwxc{} allocates memory for 23 spatial functions:
$M=20$ for the workbuffer, plus $q_0$ and its two partial derivatives.

\section{Numerical benchmarks}
\label{sect:benchmarks}

\begin{figure*}
  \includegraphics{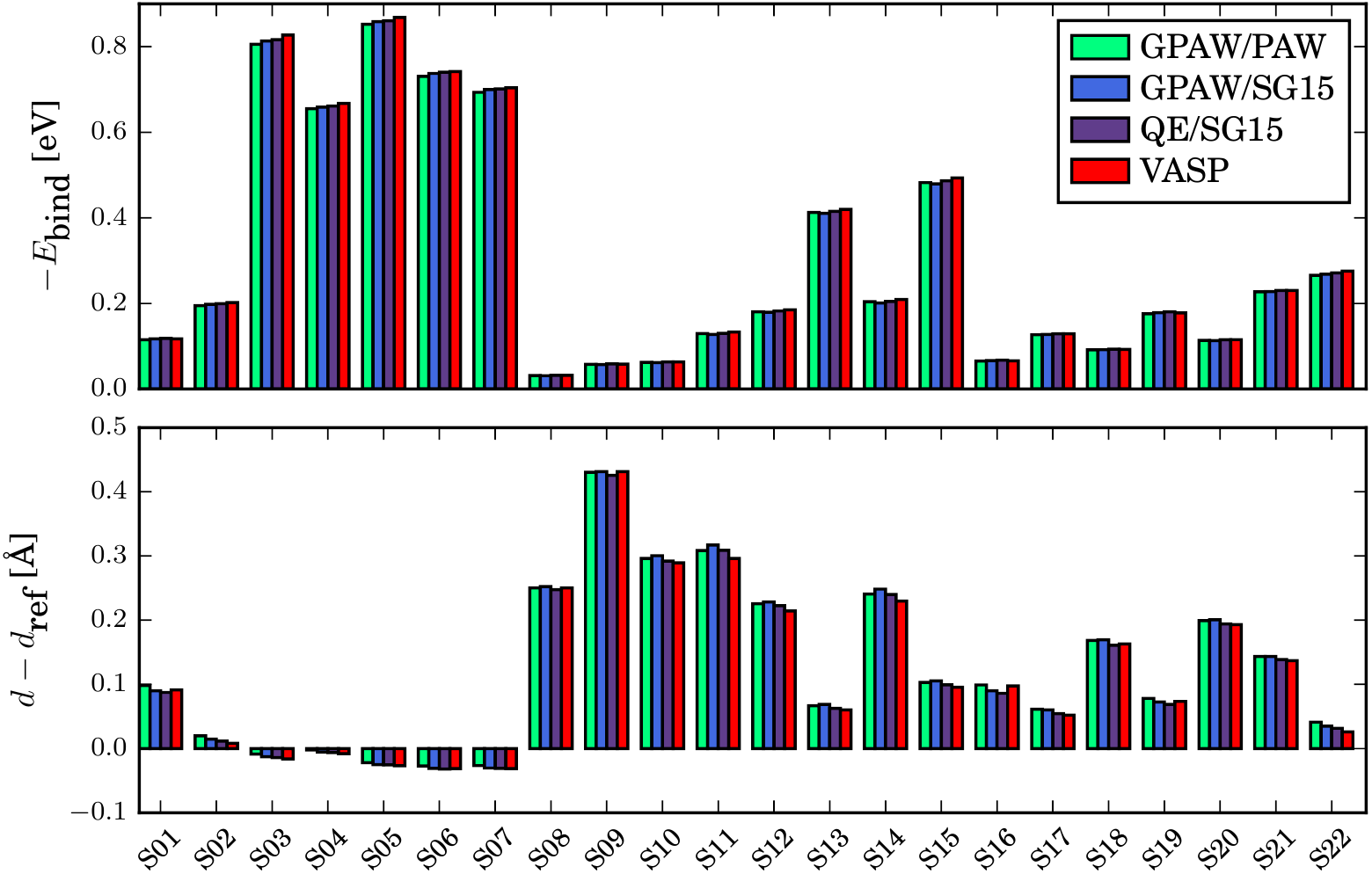}
  \caption{Benchmark of S22 dataset for different implementations of
    vdW-DF-cx.  Binding energy (top) and equilibrium bond length
    (bottom) calculated using: \gpaw{} with libvdwxc and PBE PAW datasets
    (green); \gpaw{} with \libvdwxc{} and the norm-conserving SG15
    pseudopotentials (blue); \qe{} (purple) and \vasp{} (red)
    codes using their internal vdW implementations.}
  \label{fig:s22-bars}
\end{figure*}

\noindent
To establish the numerical accuracy of \libvdwxc, the current
implementation is here benchmarked using
\gpaw{}~\cite{EnkRosMor10, MorHanJac05}, comparing to
\quantumespresso{}~\cite{GiaBarBon09} (QE) and \vasp{}~\cite{KreFur96a, KreFur96b}.
To this end, we consider the
standard S22 test set for dispersive interactions~\cite{JurSpoCer06} and
use the vdW-DF-cx functional~\cite{BerHyl14}.
Each member of the S22 test set
corresponds to a weakly bound pair of small molecules at different
intermolecular distances.  The calculated total energies therefore
yield an intermolecular binding curve.

All calculations are carried out using PAW setups or pseudopotentials
for the PBE functional.  Energies are
evaluated for a series of distances $d_i=d_{\mathrm{ref}} + i \cdot 0.025$ Å,
where $d_{\mathrm{ref}}$ is
the reference equilibrium distance~\cite{JurSpoCer06} for that S22 member;
binding energies and distances are then fitted from the 5 values of $i$
that surround the calculated minimum.
The simulation cell is such that no atom
would be closer to any edge than $10$ Å at intermolecular
separation $d^{\mathrm{ref}} + 5$ Å.
For each S22 member, the full series uses
the same simulation cell.

For each separation $d_i$, the binding energy $E[AB] - E[A] - E[B]$
is evaluated by performing the calculations for each
isolated molecule $A$ and $B$
with the atoms kept at the \emph{same} absolute positions as in
the dimer calculation $AB$.
The purpose of this is to minimize the egg-box
effect.\footnote{
  The egg-box effect arises since space is
  represented by a discrete mesh. This causes numerical noise which
  has the same periodicity as the grid and may resemble the
  shape of an egg-box. If the egg-box effect did not exist, an entire
  binding curve could be generated from one series of dimer calculations
  and \emph{two} single-molecule reference calculations.
  Instead we must do a \emph{series} of reference calculations following
  the movement of the dimer constituents.
  Note that even planewave
  codes exhibit a small egg-box effect, because typically the density
  is represented on a uniform real-space grid when evaluating the XC
  contribution.
}

%defined such that no atom was closer than
%10~Å to any cell edge for the largest separation.

%on a five-point fit with distances $d_0$, $d_0 \pm 0.025 Å$,
%and $d_0 \pm 0.05 Å$ where $d_0$ is 

\gpaw{} calculations use real-space (FD) mode with the
the standard PAW datasets (version 0.9.11271) and grid spacing 0.14~Å,
as well as the norm-conserving SG15 pseudopotentials \cite{Ham13}
with grid spacing 0.1~Å.\footnote{To completely converge the PAW
  and SG15 results it was necessary to change the finite-difference
  stencil used to evaluate $\sigma(\vec r)$ from 1st to 2nd order;
  this required small changes to the GPAW source code which are not yet released.
}
%Grid spacings are rounded slightly to match the simulation box.
The calculations use the FFT Poisson solver.

QE calculations use the norm-conserving SG15 pseudopotentials
with a planewave cutoff of 1800~eV, while VASP
calculations use the standard PAW datasets and a cutoff of 680~eV.

Figure~\ref{fig:s22-bars} shows the results: The codes produce
progressively stronger binding energies in the order \gpaw/PAW (weakest
binding), \gpaw/SG15 (+1.4 meV on average with respect to \gpaw/GPAW),
\qe{} (+3.8 meV), and \vasp{} (+6.2 meV), and as binding energies become
stronger, equilibrium bond
lengths usually become smaller
(on average 0, -0.92, -5.51, and -6.76, respectively, times $10^{-3}$ Å).

All these methods yield slightly different results,
but no single method is in strong disagreement compared to the rest.
Comparing \gpaw/PAW and \gpaw/SG15, the main difference is the representation
of the atoms.  Between \gpaw/SG15 and \qe, the main difference is real-space
versus planewave representation, although there are clearly many other
implementation differences.  Finally \vasp{} again uses PAW and has
its own vdW-DF kernel.

%The results are well converged with respect to the main computational parameters
%of each code, so the remaining differences must be ascribed to
%different PAW datasets and pseudopotentials, between real-space grids and
%planewaves (such as finite-difference stencils),
%and other computational details that cannot easily be
%unified.\footnote{Because of self-consistency, \emph{all} calculated qualities
%  have small variations: total energies, XC energies, vdW energies, etc.  However, only the total energy is
%  variational.  The variations of the individual energy components are not, and depend more strongly on pseudopotentials
%and other calculation specifics.}

%The two most technically comparable calculations are \gpaw/PAW and
%\gpaw/SG15.

Aside from the differences in atomic representation itself,
the \gpaw/PAW and \gpaw/SG15
calculations have a more profound reason to differ:
%Some difference between \gpaw/PAW and \gpaw/SG15 must be
%expected because of the very different atomic representations,
%but also for more profound reasons.
(1) the vdW-DF implementation in GPAW does not take
into account that the PAW datasets are
not norm-conserving, and (2) the calculations lack PAW corrections for
non-local XC contributions because the standard equations are derived for
semilocal functionals~\cite{Blo94}.

% energies/norms of states in generated 	extsc{Gpaw} setups (gpaw-setup <symbol>):
%H 1s                       0.988889
%B 2s(2) :    -0.344821     1.005361
%B 2p(1) :    -0.136481     0.982498
%C 2s(2) :    -0.501207     1.045313
%C 2p(2) :    -0.198967     0.954320
%N 2s(2) :    -0.676924     1.071713
%N 2p(3) :    -0.265967     0.952074
%O 2s(2) :    -0.872866     1.079797
%O 2p(4) :    -0.337925     0.876472
To elaborate on point (1), when the states are not norm-conserving,
the contribution from each state to the valence electron density does
not integrate exactly to one electron. As a result, the total pseudodensity
from which the XC energy is evaluated is
``deficient'',
and the PAW corrections only compensate for this in the semilocal terms.
This must be assumed
to spuriously affect the non-local energy.  For the chemical
species included in the S22 test set, the particular PAW datasets used
by \gpaw{} are, however, rather close to norm-conserving.  The s and p
valence states of H, B, C, N, and O have norms between 0.88 and
1.07.  This may explain that the error is small and does not
cause a clear discrepancy.

The regions far away from the atoms are also likely to provide
most of the contribution to the bonding \cite{rationalevdwdfBlugel12,BeHy13,BerHyl14,HylBerSch14}.
In these regions the density is equal to the true (all-electron) density,
and the errors mentioned do not apply.
This is another reason why the errors may be (almost) neglected.
Overall, the error
associated with the use of PAW without non-local XC corrections for
lack of norm-conservation is therefore not much larger than the
implementation error present between different codes. Note, however,
that this conclusion applies only to molecules similar
to those in the S22 set.  In particular the situation is
different for metals, for which the PAW norm is often only
around 0.3 electrons per d state.

%The \gpaw/SG15 calculations are norm-conserving and therefore do not
%have any of these errors.  These calculations are therefore in principle
%more accurate, except for the pseudopotential error.
%The QE calculations use the same kernel as \libvdwxc{}
%and use the SG15 pseudopotentials as well.
%They should in principle
%yield very similar results, but some differences remain,
%which are most likely due to
%differences between planewave and real-space methods.

In conclusion, 1) while the different methods and codes produce different
results, there are no gross outliers; and 2) the neglection
of PAW corrections has not caused the PAW calculations to particularly
disagree with pseudopotential calculations.

%As discussed, we expect the
%\gpaw/PAW results to be less accurate than the other methods due to the
%lack of vdW PAW corrections, while the \gpaw/SG15 calculations are
%nominally correct.

%\begin{table}
%  \begin{tabular}{cccc}
%    Code & GPAW/PAW & GPAW/SG15 & QE/SG15 & VASP \\
%    Atoms & PAW & SG15 & SG15 / PAW\\
%    Basis & Real-space & Real-space & PW & PW \\
%    Grid/cutoff & h=0.14 Å & h=0.1 Å & 1800 eV & 1800 eV\\
%    Cutoff & --- & --- &
%  \end{tabular}
%  \caption{}
%\end{table}

\section{Performance}
\label{sect:performance}

\noindent
The Román-Pérez--Soler algorithm solves the fundamental serial
performance issue of vdW functionals,
but parallel scalability remains essential for modern parallel computation.
In this section we investigate the scalability of \libvdwxc{}
for large systems using the DFT code \gpaw{}~\cite{EnkRosMor10,
  MorHanJac05} in the ASE framework~\cite{10.1088/1361-648X/aa680e}.

\begin{figure}
  \centering
  \includegraphics[width=8.5cm]{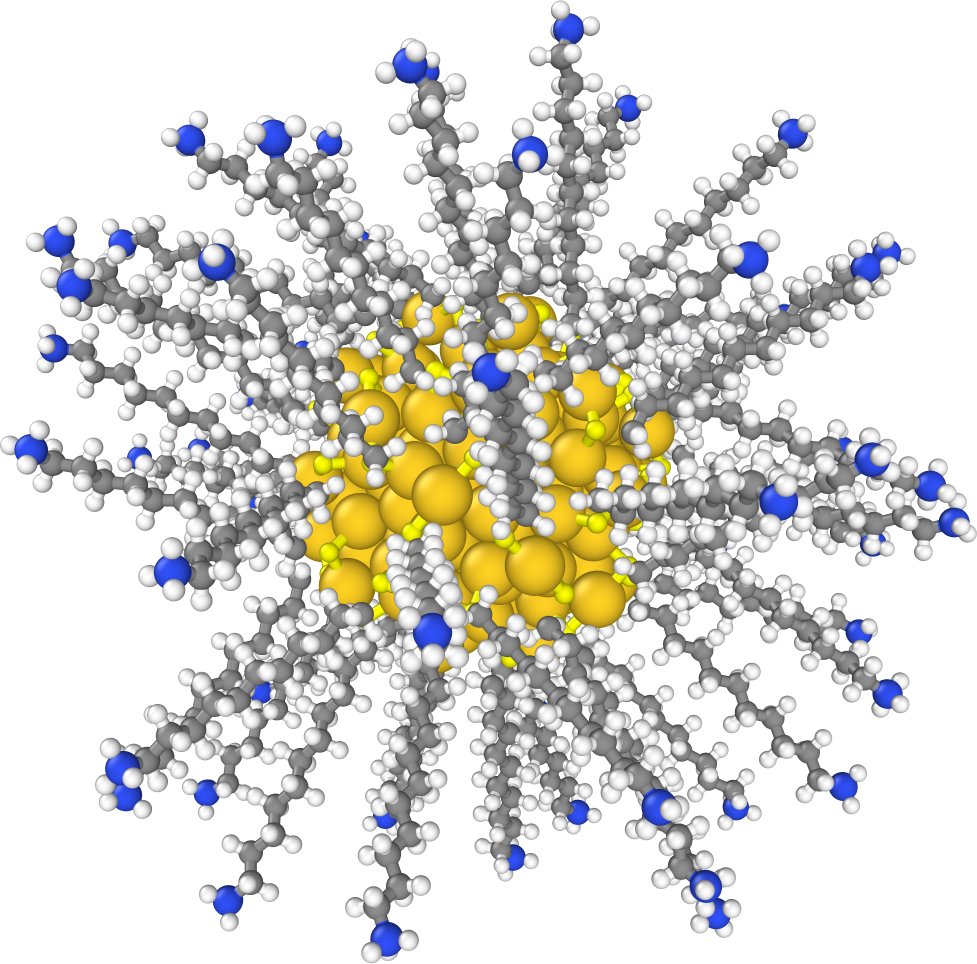}
  \caption{
    Ligand-protected Au nanocluster~\cite{HeiGurMar12} composed of
    2424 atoms with 6384 valence electrons used for testing parallel
    performance.  Rendered using \textsc{ovito}~\cite{Stu10}.
  }
  \label{fig:blasphemous-monstrosity}
\end{figure}

We consider a system of 2424 atoms:
An Au$_{144}$ nanoparticle protected by 60 extended thiol
ligands~\cite{HeiGurMar12} (SC$_{11}$NH$_{25}$) as shown in
Fig.~\ref{fig:blasphemous-monstrosity}.  The system is chosen for its
prodigious size (compared to typical DFT calculations) and because
vdW interactions are important due to the length of
ligands~\cite{LofGroMot16}.
Replicating the system up to four times (9696
atoms) along the $x$ axis enables one to generate test systems of
different size.

It is perfectly possible to test the performance of \libvdwxc{}
without doing a full DFT calculation, by simply providing an array of
arbitrary numbers. A full DFT calculation is, however, more
representative and complete,
taking into account the time spent redistributing data into the FFTW
MPI layout.

In practice the $\mathcal O(N^3)$ diagonalization of the Kohn--Sham
system will always dominate execution time for a system of this size,
except if some part of the calculation does not parallelize well or is
otherwise unnecessarily wasteful.  The primary objective of
performance testing is thus to rule out that the vdW implementation in
any way limits parallel scalability.

The calculation uses a linear combination of atomic orbitals
(LCAO)~\cite{LarVanMor09} with a single-$\zeta$ (sz) basis set for H and
single-$\zeta$ polarized (szp) basis sets for N, S and C (the
ligands), which are smaller than the usually recommended
double-$\zeta$ polarized (dzp) basis set. The new and improved ``p-valence''
basis set~\cite{KuiSakRos15} is used for Au. All timings are averaged
over 15 iterations of the self-consistency cycle.
The single nanoparticle test system has $320^3$ points in real-space (density and
potential including vdW XC are evaluated on a $640^3$ grid), 11112
atomic orbitals, and 6384 valence electrons within 3352 electronic
states. The simulation box volume is (58 Å)$^3$.

The test environment was the Niflheim supercomputer at the Technical
University of Denmark. Each test node has two Intel Ivy Bridge Xeon
E5-2650 v2 8-core 2.6 GHz CPUs (16 cores per node) using quad data
rate Infiniband interconnect.

\begin{figure*}
  \includegraphics{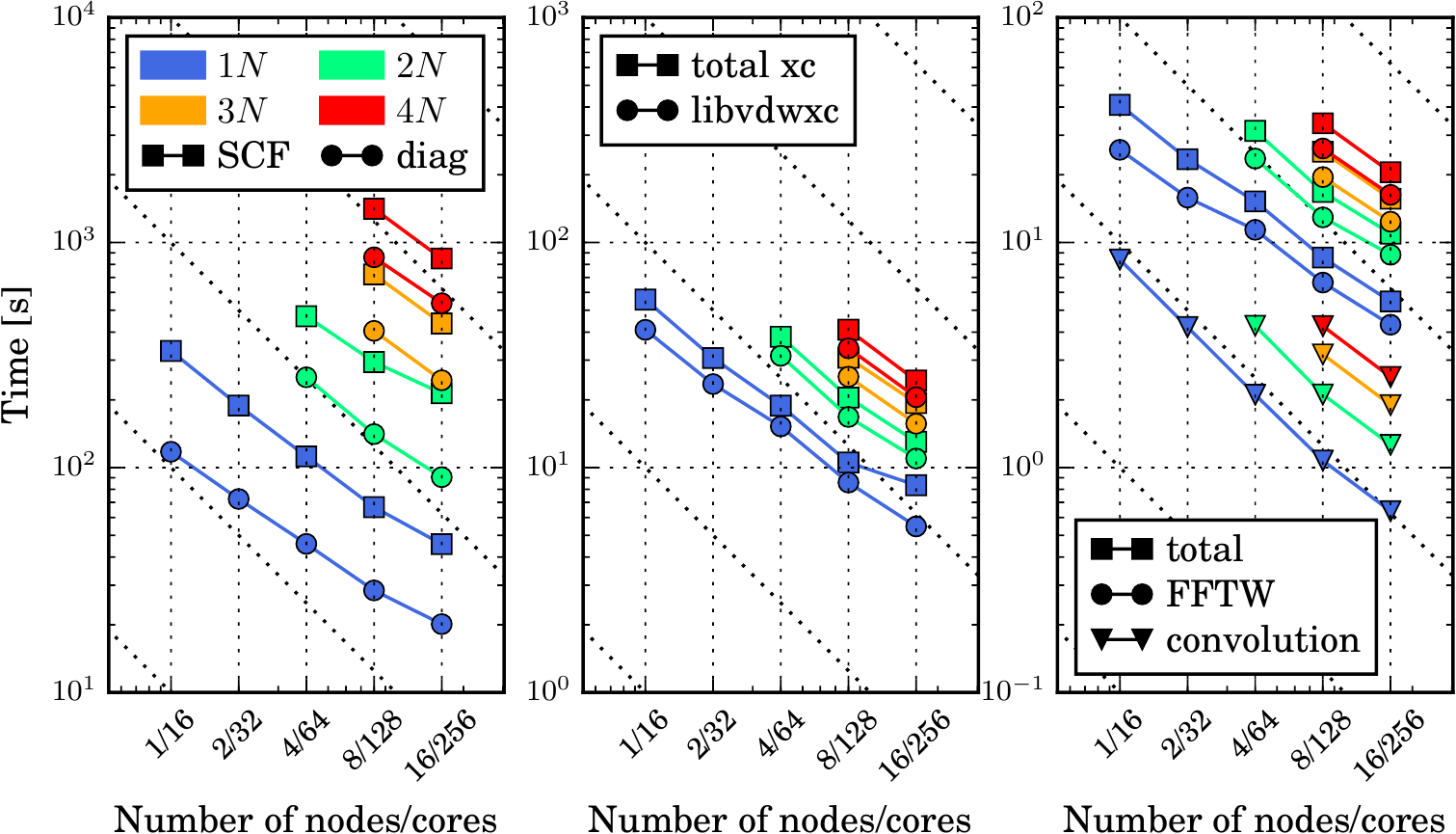}
  \caption{Scaling benchmarks for \libvdwxc.
    Left: Time per full self-consistent field (SCF) step in
    DFT calculation (squares) as a
    function of number of compute nodes, each with 16 cores,
    calculated for 1 (blue), 2 (green), 3 (orange) and 4 (red)
    repetitions of the nanocluster test system (see
    Fig.~\ref{fig:blasphemous-monstrosity}). Also shown is the time
    hereof
    spent diagonalizing the Hamiltonian (circles), which is the most
    expensive part of the SCF step.  Dotted lines indicate perfect
    parallel scaling.
    Middle: Timing spent for full XC calculation (\libvdwxc{} +
    semilocal + distribution of data) (squares) and, hereof, time
    spent in \libvdwxc{} (circles).
    Right: Time spent in \libvdwxc{} (squares) and, hereof, time spent
    doing FFTs and convolution. The remaining operations are
    inexpensive.
  }
  \label{fig:scaling}
\end{figure*}

Figure~\ref{fig:scaling} shows the scaling of large calculations based
on the nanoparticle system described above. Since there is no single
natural choice for the main \gpaw{} parallelization parameters (ScaLAPACK
layout, domain decomposition and band parallelization), these
parameters were chosen on a best-effort basis for each system.  As a
result, the timings shown in the left-most figure ought be considered
only representative. The \libvdwxc{} timings do not depend on those
parameters, however, so timings in the remaining figures are
consistent.  \libvdwxc{} uses the standard FFTW-MPI
distribution with the (automatically chosen) blocksize that most evenly
divides the number of
grid points.  Overall the \libvdwxc{} calculation performs in the
strong scaling limit at least as well as the $\mathcal O(N^3)$
diagonalization.  Note that for simplicity we have benchmarked only
with FFTW-MPI (not the more scalable PFFT).

Calculations normally use the larger dzp basis sets.
This would make the vdW part of the calculation less expensive in
comparison; the test therefore represents a worst-case comparison of
vdW performance with respect to the overall DFT calculation.
Real-space or planewave calculations of a system of this size would be
even more expensive.

\section{The library} \label{sec:thelibrary}

\noindent \libvdwxc{} is written in C and uses
the standard GNU build system, autotools, for compilation.
Software requirements are the FFTW3 library,
an MPI library (for parallelism), and optionally the PFFT library.
These can be specified at compilation time.
The installation produces the main library (\texttt{libvdwxc.so} or
\texttt{libvdwxc.a}) plus Fortran bindings.

A calculation comprises the following steps:
\begin{enumerate}
\item Call \texttt{vdwxc\_new} to create an empty \texttt{vdwxc\_data}
  data structure for a given vdW functional.
  The \texttt{vdwxc\_data} structure contains the internal data associated with
  a calculation and is always passed as a handle to libvdwxc functions.
\item Call \texttt{vdwxc\_set\_unit\_cell} to specify the number of
  grid points in each direction as well as the unit cell.
\item Call one of \texttt{vdwxc\_init\_serial},
  \texttt{vdwxc\_init\_mpi}, or \texttt{vdwxc\_init\_pfft}
  to initialize the FFTW backend and allocate the necessary memory.
\item Call \texttt{vdwxc\_calculate} any number of times, passing
  pointers to arrays with the input density and output potential.
\item Call \texttt{vdwxc\_finalize} to deallocate the memory.
\end{enumerate}
All calculations use double precision floating point numbers.
Further convenience functions are provided, e.g., to print the state of a
\texttt{vdwxc\_data} data structure, or to check which of the
optional libraries are available at runtime.

The code below illustrates the simplest possible form of a program using \libvdwxc:
\begin{lstlisting}
#include <stdio.h>
#include <vdwxc.h>

void main()
{
    double rho[] = {1.0, 2.0, 3.0, 4.0}; // input: density & gradient
    double sigma[] = {0.5, 1.5, 2.5, 3.5};
    double dedrho[] = {0.0, 0.0, 0.0, 0.0}; // output: derivatives
    double dedsigma[] = {0.0, 0.0, 0.0, 0.0};

    vdwxc_data vdw = vdwxc_new(VDWXC_DF1); // vdW-DF1 functional
    vdwxc_set_unit_cell(vdw, 1, 2, 2, // 1x2x2 grid; then cell vectors
                        1.0, 0.0, 0.0, 0.0, 2.0, 0.0, 0.0, 0.0, 2.0);
    vdwxc_init_serial(vdw);
    double energy = vdwxc_calculate(vdw, rho, sigma, dedrho, dedsigma);
    vdwxc_finalize(&vdw);
    printf("Energy %f Hartree\n", energy);
}
\end{lstlisting}

The greatest challenge facing a DFT code developer who wants to call
\libvdwxc{}, is to redistribute the density into the correct parallel
format as described in Section~\ref{sect:parallelization}.
This is difficult to generalize since DFT codes use different
parallelizations, but would
typically involve a call to \texttt{MPI\_Alltoallv} after
establishing suitable buffers.
Nevertheless DFT codes tend to provide tools
for accomplishing this task.
We implemented the redistribution in \gpaw\
using the existing parallel Python framework.
\octopus\ already
happened to support redistributions of the required type since it uses the
same FFT libraries.

%Parallel redistributions are notoriously difficult, and cannot be
%implemented once and for all because they depend on what parallel
%format the calling DFT code already uses.  Yet many DFT codes already
%perform redistributions

\libvdwxc{} is free software distributed under the GNU General Public
License 3 or any later
version.\footnote{
  We expect that \libvdwxc{} will be of most interest to codes that
  already link to \libxc.  The vast majority of these codes is released
  under GPL; since FFTW and PFFT are released under GPL as well, GPL
  is a logical choice for \libvdwxc.  In the future
  we may consider making it possible to use the library
  from non-GPL codes ---such as \libxc\ which is released under LGPL---
  if this benefits the community.}
Source code and documentation are available from the
homepage~\cite{libvdwxc-homepage}
including the current stable release, \libvdwxc{} 0.2.0.
Development takes place openly on GitLab~\cite{libvdwxc-dev}.

\section{Concluding remarks}

\noindent \libvdwxc{} enables the efficient computation of the non-local
correlation energies and potentials of the vdW-DF
functionals on massively parallel architectures. The energies are
evaluated using the vdW kernel parametrization from
\qe{}~\cite{GiaBarBon09}, while support for the \gpaw{} kernel
format~\cite{Wellendorff2010} and others is planned.
Furthermore, extending the library to support spin polarized calculations
 \cite{Thonhauser_2015:spin_signature} is underway.

Up to now, \libvdwxc{} has therefore been interfaced with the
\gpaw{} and \octopus{} codes,
which are attractive
starting points for different reasons. \gpaw{} features an
extremely efficient linear combination of atomic orbitals
implementation for DFT~\cite{LarVanMor09} (as demonstrated
in Section \ref{sect:performance}) and time-dependent
DFT~\cite{KuiSakRos15}.
%The  has been demonstrated
%here by calculations for a 4\,nm diameter ligand-protected gold
%nanoparticle~\cite{HeiGurMar12}.
\octopus{}, on the other hand,
has been used to simulate coherent charge transfer on very
large supermolecular triads with non-adiabatic
dynamics~\cite{RozFalSpa13}.
Since vdW forces appear to play a role in the
charge separation in organic photovoltaics~\cite{KarCraKui15}, vdW-DF
functionals in \octopus{} are particularly appealing.
Furthermore an interface for the planewave PAW code
\textsc{Abinit}~\cite{Gonze20092582} is planned.

%Currently the \gpaw{} and \octopus{} codes can be used with \libvdwxc.
The non-local vdW functionals can be combined with the semilocal functionals
from \libxc{}~\cite{MarOliBur12}
to form various other functionals of the vdW-DF family.  Hence
vdW-DF-optPBE, vdW-DF-optB88, vdW-DF-C09, vdW-DF-BEEF,
and vdW-DF-mBEEF are all
available in \gpaw{} with \libvdwxc.

Tests of vdW-DF-cx on the S22 set of molecular binding curves show that
binding energies and bond lengths
differ somewhat across several tested vdW implementations,
as is commonly the case.
The deviations are, however, rather small
for the tested implementations.
%\libvdwxc{} is parallelized using MPI and FFTW, and optionally PFFT.
From tests of up to 10000 atoms, we conclude that the parallel
scalability of \libvdwxc{} should be sufficient for any kind of system
accessible to ordinary Kohn--Sham DFT.

%In summary, \libvdwxc{} enables the efficient and accurate calculation
%of the non-local correlation energy for XC functionals in the vdW-DF
%family.
%Interfaces for \gpaw{} and \octopus{} are already available, while
%additional interfaces will be added in the future.
We expect this
library to enhance the widespread use of the vdW-DF
method. We hope that the availability of an implementation
that is independent from any DFT code will lead to more reliable and reproducible
calculations of the small energy differences that are characteristic
of vdW interactions, and that
\libvdwxc{} can serve as a
common testing ground for new developments within the framework of
van der Waals functionals and related methods.
Indeed the spline decomposition algorithm could
be adapted to entirely different non-local functionals
if they are expressible as a convolution.
What happens will depend on the community since \libvdwxc{} is open
to contributions.

%Development version and stable releases (currently only 0.2.0)
%are available from GitLab~\cite{libvdwxc-dev,libvdwxc-stable}.
%Documentation is hosted at Read the Docs~\cite{libvdwxc-readthedocs}.

%\begin{acknowledgments}
\subsection*{Acknowledgements}
\noindent
A.\,H.\,L.\ acknowledges funding from the European Union's Horizon 2020
research and innovation program under grant agreement no.~676580 with
The Novel Materials Discovery (NOMAD) Laboratory, a European Center of
Excellence.
M.\,K.\ is grateful for Academy of Finland Postdoctoral Researcher
funding under Project No.~295602.
This work has been supported by the Swedish Research
council (VR), the Knut and Alice Wallenberg foundation, the Swedish
Foundation for Strategic Research, and the Chalmers Area of Advance
Materials Science.
We acknowledge access to the supercomputer Niflheim at the Department
of Physics, Technical University of Denmark.
A.\,H.\,L.\ and Y.\,P.\ thank A.\,L.~Pacas for stimulating discussions.
The authors thank Elena (Heikkilä) Kuisma for
providing the test system used for the parallel benchmarks,
and Kristian Berland and Carlos de Armas for further discussions.
%\end{acknowledgments}

\appendix*

\section{$q_0(\vec r)$}
\noindent
Dion \emph{et al.}~\cite{DioRydSch04} introduced the quantity
$q_0$ as a modification of the local Fermi wave vector,
\begin{align}
q_0 = \frac{\epsilon_{\XC}^0}{\epsilon_\X^{\mathrm{LDA}}} k_F,
\end{align}
where $k_F=(3 \pi^2 n)^{1/3}$ is the local Fermi wavevector and
$\epsilon_{\XC}^0$ is the LDA energy per particle modified by a
a gradient correction describing screened exchange.
For spin paired system, the
full expression can be written as \cite{PhysRevB.76.125112}
\begin{align}
q_0 = k_F - \frac{4\pi}{3} \epsilon_\cc^{\mathrm{LDA}}
    - \frac{Z_{ab}}{36}\frac{\sigma}{k_F n^2}.
\end{align}
The LDA correlation energy per particle is evaluated using the
parametrization of Perdew and Wang~\cite{PerWan92},
\begin{align}
\epsilon_\cc^{\mathrm{LDA}}%^{\mathrm{LDA}}
= -2 A(1 + \alpha_1 r_s) \log \left[
  1 + \frac{1}{2 A \sum_{i=1}^4 \beta_i r_s^{i/2}}
  \right].
\end{align}
The constants are $A=0.031091$, $\alpha_1=0.2137$, $\beta_1=7.5957$,
$\beta_2=3.5876$, $\beta_3=1.6382$, $\beta_4=0.49294$, and
$r_s$ is the Wigner--Seitz radius, $r_s^3 = 4 \pi n / 3$.
vdW-DF1 uses $Z_{ab}=-0.8491$ whereas vdW-DF2
uses $Z_{ab}=-1.887$.  Other functionals in the vdW-DF
family generally use one of these two values,
and this is the only common variation of the
non-local correlation energy.

%Done.........

%$q_0$ is given by
%\begin{align}
%  q_0(\vec r) &= k_F + A_1(\vec r)
%  \log\left(1 + \frac{1}{A_2(\vec r)}\right)
  %\log(1 + 1/A_2(\vec r))
%  - D(\vec r),
%\end{align}
%where
%\begin{align}
%  A_1(\vec r) &= \frac83 \pi A (1 + a_1 r_s),\\
%  A_2(\vec r) &= 2 A (b_1 r_s^{1/2} + b_2 r_s +
%  b_3 r_s^{3/2} + b_4 r_s^2),\\
%  D(\vec r) &= \frac{Z_{ab}}{36 k_F^2} \frac{\sigma(\vec r)}{ n(\vec r)^2},\\
%  r_s &= \left(\frac{3}{4 \pi n(\vec r)}\right)^{1/3},\\
%  k_F &= (3 \pi^2 n(\vec r))^{1/3}.
%  %\sigma(\vec r) &= \norm{ \nabla n(\vec r)}^2.
%\end{align}
%The constants are $A=0.031091$, $a_1=0.2137$, $b_1=7.5957$,
%$b_2=3.5876$, $b_3=1.6382$, $b_4=0.49294$.  These values are from
%the Perdew--Wang 1992 parametrization of LDA correlation
%.

For calculations that use pseudopotentials or PAW, the density is
generally relatively smooth and values of $q_0$ are generally smaller
than 5 atomic units.  Therefore the $q_0$ grid ends at 5 a.u.,
and to retain good numerical behavior towards this limit, the values of $q_0$
are filtered through the saturation function from Ref.~\cite{RomSol09}
given by
\begin{align}
  h_a(x) = a
  \left[1 - \exp\left(
    -\sum_{m=1}^{12} \frac1m \left[\frac x a \right]^m
    \right)\right].
\end{align}
This function has the property that $h_a(x) \approx x$ for small $x$,
and $h_a(x) \rightarrow a$ from below for $x \rightarrow \infty$.
Thus, in actual computations, $q_0(\vec r) = h_{a=5}(q_0^{\mathrm{orig}}(\vec r))$.

%The derivatives of $q_0$ are probably too tedious to list here.
% PE: ... and unnecessary

%\bibliographystyle{apsrev4-1} % Tell bibtex which bibliography style to use
%\bibliographystyle{abbrv}
\section*{References}
\bibliographystyle{unsrt}
\bibliography{lit}

\end{document}